\documentclass[notoc]{JHEP3}

\usepackage{epsfig}

\newcommand{\beq}{\begin{equation}}
\newcommand{\eeq}{\end{equation}}
\newcommand{\bea}{\begin{eqnarray}}
\newcommand{\eea}{\end{eqnarray}}
\newcommand{\nn}{\nonumber}

\def\eqn#1{Eq.~(\ref{#1})}
\def\eqns#1#2{Eqs.~(\ref{#1}) and~(\ref{#2})}

\def\sec#1{Section~{\ref{#1}}}

\newcommand\fverb{\setbox\pippobox=\hbox\bgroup\verb}
\newcommand\fverbdo{\egroup\medskip\noindent%
                        \fbox{\unhbox\pippobox}\ }
\newcommand\fverbit{\egroup\item[\fbox{\unhbox\pippobox}]}
\newbox\pippobox

\def\ord{{\cal O} }
\def\cM{{\cal M}}
\def\eps{\epsilon}
\def\ione{{\bf I}^{(1)}}
\def\itwo{{\bf I}^{(2)}}
\def\htwo{{\bf H}^{(2)}}

\def\MS{$\overline{\rm MS}$}
\def\cg{c_\Gamma}
\newcommand\sss{\scriptscriptstyle}
\newcommand\as{\alpha_{\sss S}}
\newcommand\gs{g_{\sss S}}
\def\tgs{\tilde\gs}

\def\CF{C_{\sss F}}
\def\CA{C_{\sss A}}
\def\NF{N_{\sss F}}

\title{The quark Regge trajectory at two loops}

\author{A.~V.~Bogdan\\
Budker Institute for Nuclear Physics,
Novosibirsk State University\\ 630090 Novosibirsk, Russia\\
E-mail: \email{bezov@ngs.ru}}

\author{V.~Del~Duca\\
Istituto Nazionale di Fisica Nucleare, Sez. di Torino\\
via P. Giuria, 1 - 10125 Torino, Italy\\
        E-mail: \email{delduca@to.infn.it}}

\author{V.~S.~Fadin\\
Budker Institute for Nuclear Physics,
Novosibirsk State University\\ 630090 Novosibirsk, Russia\\
E-mail: \email{V.S.Fadin@inp.nsk.su}}

\author{E.~W.~N.~Glover\\
Institute for Particle Physics Phenomenology,
University of Durham\\ Durham, DH1 3LE, U.K.\\
E-mail: \email{E.W.N.Glover@durham.ac.uk}}

\abstract{
The hypothesis of the quark Reggeization is tested by taking the high-energy
limit of the two-loop amplitude for quark-gluon scattering which was
recently calculated. The limit is compatible with the Reggeization in the
leading and the next-to-leading orders and allows the determination of the
quark trajectory   in the two-loop approximation. The trajectory is presented
as an expansion in powers of $(D-4)$ for the space-time dimension $D$ tending
to the physical value $D=4$.
}

\preprint{{~DCTP/02/08},{~IPPP/02/04},{~DFTT 03/2002},{Budker INP 2002-4}{~hep-ph/0201240}}

\begin{document}

\section{Introduction}
\label{sec:a}

In the limit of large center of mass energy $\sqrt s$ and fixed
momentum transfer $\sqrt {-t}$  the most appropriate approach for
the description of the scattering amplitudes is given by the
theory of the complex angular momenta or Gribov-Regge theory. One
of the remarkable properties of QCD is the Reggeization of its
elementary particles. Unlike QED, where the electron Reggeizes
\cite{gellmann,McCoy:1976ff}, but the photon remains elementary
\cite{mandel}, in QCD both the gluon
\cite{Grisaru:1973vw,Grisaru:1973cf,Lipatov:1976zz,Fadin:1975cb,Kuraev:1976ge,Kuraev:1977fs,Balitsky:1978ic}
and quark \cite{Fadin:1977jr} Reggeize. We use here the notion
``Reggeization" in the strong sense \cite{Fadin:1998fv}, that
means not only the existence of a Reggeon with corresponding quantum
numbers (including signature) and trajectory, but as well the
dominance of the Reggeon contribution to the amplitudes of the
processes with these quantum numbers in each order of perturbation
theory. For example, parton-parton scattering amplitudes  in QCD
are dominated by gluon exchange in the crossed channel. The
Reggeized gluon is a colour octet state of negative signature in
the $t$-channel, {\it i.e.} odd under  $s \leftrightarrow u$
exchange, and to leading logarithmic (LL) accuracy in $\ln(s/|t|)$
the virtual radiative corrections to the parton-parton scattering
amplitude with the colour octet state and negative signature can
be obtained,  to all orders in $\as$, by the
replacement~\cite{Kuraev:1976ge}
\begin{equation}
{s\over t} \to {1\over 2}\left[
  \left({-s\over -t}\right)^{j_G(t)}
- \left({s\over -t}\right)^{j_G(t)} \right] \, ,\label{sud}
\end{equation}
where  $j_G(t)\equiv \omega(t)+1$ is called the Regge trajectory of the gluon.

The property of the Reggeization is very important for  high energy QCD. The
BFKL equation~\cite{Fadin:1975cb,Kuraev:1976ge,Kuraev:1977fs,Balitsky:1978ic}
for the resummation of the leading logarithmic radiative corrections to scattering
amplitudes for processes with gluon exchanges in the $t$-channel is based on
the gluon Reggeization. The Pomeron, which determines the high energy behaviour
of cross sections, and the Odderon, responsible for the difference of particle
and antiparticle cross sections, appears in QCD as a compound state of two and
three Reggeized gluons respectively. Similarly colorless objects constructed
from Reggeized quarks and antiquarks should also be relevant to the
phenomenological description of processes involving the exchange of quantum
numbers.

The basic parameters of the Reggeons are their trajectories and the interaction
vertices.  To leading logarithmic accuracy, $\omega(t)$ is related to a
one-loop transverse-momentum  integration. In dimensional regularization in
$D=4+2\epsilon$ dimensions this can be written as
\begin{equation}
\omega(t) = \gs^2\,
\left(\mu^2\over -t\right)^{-\epsilon} c_{\Gamma}\,\omega^{(1)} + {\cal
O}(\gs^4) ,\label{alph}
\end{equation}
with
\begin{eqnarray}
\omega^{(1)} &=& -\CA{2 \over \epsilon}\, ,\label{omega1}\\
c_{\Gamma} &=& {1\over (4\pi)^{2+\epsilon}}\, {\Gamma(1-\epsilon)\,
\Gamma^2(1+\epsilon)\over \Gamma(1+2\epsilon)}\, ,\label{cgam}
\end{eqnarray}
and $\CA = N$.
The resummation of the real and virtual
radiative corrections to parton-parton scattering  amplitudes with gluon
exchange in the crossed channel is related by unitarity to the imaginary part
of the elastic amplitudes with all possible colour exchanges in the crossed channel.
The radiative corrections to these amplitudes
are resummed through the BFKL equation i.e. a
two-dimensional integral equation which describes
the interaction of two Reggeized gluons in the crossed channel.

The integral equation is obtained to LL by computing the one-loop
corrections to the gluon exchange in the $t$ channel.  They are
formed by a real correction: the emission of a gluon along the
ladder~\cite{Lipatov:1976zz} and a virtual correction: the
one-loop Regge trajectory, \eqn{sud}. The BFKL equation is then
obtained by iterating recursively these one-loop corrections to
all orders in $\alpha_s$, to LL accuracy.

In recent years, the BFKL equation has been improved to next-to-leading
logarithmic (NLL) accuracy~\cite{Fadin:1998py,Camici:1997ij,Ciafaloni:1998gs}.
A necessary ingredient has been the calculation of the two-loop Regge
trajectory of the gluon~\cite{Fadin:1995xg,Fadin:1996tb,Fadin:1996km,Blumlein:1998ib}.
Recently the gluon Regge trajectory was re-evaluated \cite{DelDuca:2001gu}, in
a completely independent way, by taking the high energy limit of the two-loop
amplitudes for parton-parton scattering  with gluon exchanges in the
$t$-channel. The validity of the gluon Reggeization to NLL was confirmed and
full agreement with previous results was found.

So far attention has focussed mainly on processes dominated by
Reggeized gluon exchange.   However, let us consider a scattering
process  with fermion exchange, namely quark-gluon scattering,
which proceeds via the exchange of a quark in the crossed channel,
and let us take  the limit $s\gg |u|$. Since in the center-of-mass
frame of a two-particle scattering $u = - s (1+\cos\theta)/2$, the
limit $s\gg |u|$ corresponds to backward  scattering. By crossing
symmetry, this is equivalent to quark pair annihilation into two
gluons at small angles (in which case the roles of $s$ and $t$ are
exchanged). The contribution from the exchange of a colour triplet
state and of positive signature {\it i.e.} even under $s
\leftrightarrow t$ exchange Reggeises, so that  the virtual
radiative corrections to the quark-gluon scattering amplitude with
the colour triplet state and positive signature in the limit $s\gg
|u|$ can be obtained, to all orders in $\as$ and to LL accuracy in
$\ln(s/|u|)$, by the replacement~\cite{Fadin:1977jr}
\begin{equation}
\sqrt{s\over - u} \to {1\over 2} \;\sqrt{s\over - u} \; \left[
 \left({s\over - u}\right)^{\delta_T(u)}
 +\left({-s\over - u}\right)^{\delta_T(u)}
 \right]\, ,\label{sudfer}
\end{equation}
where the quark Regge trajectory is $j_Q(u) = \delta_T(u)+1/2$.
$\delta_T(u)$ is related to a one-loop transverse-momentum
integration, which, for massless quarks and up to replacing the colour factor
$\CA$ with $\CF$,
is the same as for the gluon trajectory. Thus in dimensional regularization
$\delta_T(u)$ can be written as,
\begin{equation}
\delta_T(u) = -\gs^2\, \CF\, {2\over\epsilon}
\left(\mu^2\over -u\right)^{-\epsilon} c_{\Gamma}\, ,\label{delt}
\end{equation}
with $\CF = (N^2-1)/(2N)$, and $c_{\Gamma}$ given in \eqn{cgam}.
This is similar to what
happens in QED, where the electron also
Reggeizes~\cite{gellmann,McCoy:1976ff}.
Analogously to the gluon case, an equation was derived~\cite{Fadin:1977jr}
to resum the real and virtual radiative corrections to exchanges
with arbitrary colour state and signature.
However, unlike from the BFKL equation for gluon exchange,
that equation has not been solved.

In this paper we explicitly take the high-energy limit of the
one-loop~\cite{Ellis:1986er,Kunszt:1994sd} and of the two-loop amplitudes
for quark-gluon scattering~\cite{Anastasiou:2001sv}. This allows us to
test the validity of the quark Reggeization
and to calculate the two-loop Regge trajectory of the quark.

Our paper is organised as follows.   In Sect.~2, we discuss the general
structure of the quark-gluon scattering amplitudes, we evaluate them at
tree and one-loop level in the limit $s\gg |u|$ and decompose them according
to the irreducible colour representations exchanged in the $u$ channel. In
Sect.~3 we discuss the Regge ansatz for resumming the LL and NLL and evaluate
the leading and next-to-leading order corrections for the
interference of the tree-amplitude with the Reggeized ansatz in terms of the
gluon trajectory and the impact factors.  Sect.~4 is devoted to the analysis of
the one- and two-loop Feynman diagram calculations in the high energy limit and
to the extraction of the LL, NLL and NNLL behaviours.  We make a detailed
comparison of the two approaches and show that the two approaches are
compatible at the LL and NLL level.   This allows the determination of the
quark trajectory to two-loop order which we give as an expansion in powers of
$(D-4)$ for the space-time dimension $D$ tending to the physical value $D=4$.
Finally, our findings are summarized in Sect.~5.

\section{The structure of the quark-gluon scattering amplitude}
Let the incoming quark and gluon have momenta $p_a$ and $p_b$
respectively, and the outgoing quark and gluon have momenta
$p_{a'}$ and $p_{b'}$ respectively.

The most general possible colour decomposition for the amplitude
${\cal M}$ is
\begin{equation}
{\cal M}  = 2(T^b T^{b'})_{a'a} \;{\cal A} +2(T^{b'}
T^{b})_{a'a}\;{\cal B} +\delta^{bb'} \delta_{a'a}\;{\cal C},
\label{eq:coldec}
\end{equation}
where $b$, $b'$ are the colours of the two gluons and $a$, $a'$
are the colours of the quarks. The factor 2 in \eqn{eq:coldec} is
due to our choice for the normalization of the fundamental
representation matrices, i.e. $tr(T^aT^b) = \delta^{ab}/2.$ The
functions ${\cal A}$, ${\cal B}$ and ${\cal C}$ are
colour-stripped sub-amplitudes where we have suppressed all
dependence on the particle polarisations and momenta. At lowest
order, for general kinematics, only the first two colour
structures are present. They are related by $t \leftrightarrow u$
exchange, thus in the $s\gg |u|$ limit only one of them
contributes. The third colour structure appears first at one-loop
order and is symmetric under $u$ and $t$ exchange.

\subsection{Quark-gluon amplitudes at tree and one-loop level}
\label{sec:twoone}

In the high-energy limit $s\gg |u|$, quark-gluon scattering (or
any other crossing symmetry related process) is dominated by quark
exchange. In this limit, at tree level accuracy only the
configuration for which the outgoing gluon has the same helicity
as the incoming quark will contribute. This is equivalent to say
that in the limit $s\gg |u|$ helicity is conserved in the $s$
channel. Thus, once the helicity along the quark line is fixed, of
the two gluon helicity configurations allowed at tree level, 
only one dominates. The only other leading helicity
configuration is the one obtained from this by parity, which flips
the helicity of all the particles in the scattering.

Using the colour decomposition (\ref{eq:coldec}) in the helicity
formalism~\cite{Mangano:1991by}, we note that 
there is only one independent colour-stripped tree sub-amplitude, the
Parke-Taylor sub-amplitude \cite{Parke:1986gb}. Using the spinor
products in the limit $s\gg |u|$ (which can be easily derived from the
amplitudes listed in the appendix of
Ref.~\cite{DelDuca:2000ha} in the limit $s\gg |t|$) 
in the sub-amplitude for the leading
helicity configuration, we obtain the 
\beq \mathcal{A}^{(0)}(p_a^-,p_b^-;p_{a'}^-,p_{b'}^-) = - i\,
\gs^2\, \sqrt{s\over -u} \left({p_{a'_\perp}\over
|p_{a'_\perp}|}\right)^2\, ,\label{suba} \eeq
where complex transverse coordinates $p_{\perp} = p^x + i p^y$
have been used, and
the superscipts in the argument on the left-hand side label
the parton helicities\footnote{Note that, contrary to the conventions
of the helicity formalism where all particles are taken as outgoing, 
here we label momenta and helicities as in the physical scattering.}. 
Using Eq.~(\ref{eq:coldec}), the amplitude
for quark-gluon scattering $q_a\, g_b\to q_{a'}\, g_{b'}$ for a
generic tree-level helicity configuration may be written as,
\bea \lefteqn{ {\cM}^{(0)}(p_a^{\nu_q},
p_b^{\nu_g};p_{a'}^{\nu_q}, p_{b'}^{\nu_{g'}}) }\nn\\ &=& - 2i
\left[\gs\, (T^b)_{a'i}
C_{qg}^{(0)}(p_a^{\nu_q},p_{b'}^{\nu_{g'}}) \right] \sqrt{s\over -
u} \left[\gs\, (T^{b'})_{i\, a}\, C_{gq}^{(0)}(p_b^{\nu_g},
p_{a'}^{\nu_{q}}) \right]\, ,\label{elas} \eea
there the $\nu$'s denote the parton helicities, and we have
explicitly enforced helicity conservation along a massless fermion
line. From \eqn{suba}, the tree-level coefficient function $C^{(0)}$ is,
\beq C^{(0)}_{q\,g}(p_a^-,p_{b'}^-) =
C^{(0)}_{g\,q}(p_b^-,p_{a'}^-) = {p_{a'_\perp}\over
|p_{a'_\perp}|}\, ,\label{zerocoeff} \eeq
while for $s \gg |u|$ the unequal helicity coefficient functions of type
$C^{(0)}(\pm\mp)$ are subleading. Squaring and summing \eqn{elas}
over helicity and colour, we obtain,
\beq \sum_{hel,\, col} |{\cM}^{(0)}|^2 = 8\ C_{\sss F}^2\ N_c\
\gs^4\ {s\over -u} = {128\over 3}\ \gs^4\ {s\over -u}\,
.\label{squaretree} \eeq %
in agreement with the $s\gg |u|$ limit of the squared tree
quark-gluon amplitudes.

In Ref.~\cite{Kunszt:1994sd}, the coefficients in the
 colour decomposition (\ref{eq:coldec}) of the one-loop amplitude
 for the quark-gluon scattering have been calculated in the 't
Hooft-Veltman and in the dimensional reduction infrared schemes.
Using the results of Ref.~\cite{Kunszt:1994sd}, the sub-amplitude of
Eq.~(\ref{suba}) and the tree amplitude of Eq.~(\ref{elas}), we can write the
unrenormalised one-loop amplitude in the helicity configuration of
\eqn{suba}, in the 't Hooft-Veltman scheme and in the limit $s\gg
|u|$ as,
\begin{eqnarray}
\lefteqn{ {\cM}^{(1)}(p_a^-,p_b^-;p_{a'}^-,p_{b'}^-)  
= }\nonumber \\
&&\tgs^2(u) {\cM}^{(0)}(p_a^-,p_b^-;p_{a'}^-,p_{b'}^-)  \left[ {2C_{\sss
F}\over {-\eps}} \ln\left({s\over -u}\right) - {2(\CA + \CF )\over
\eps^2} + {3\CF\over\eps} + (\pi^2 -7) \CF \right]\nonumber \\
&+&
 \tgs^2(u)\,\delta^{bb'}\delta_{a'a}\
 \mathcal{A}^{(0)}(p_a^-,p_b^-;p_{a'}^-,p_{b'}^-) \,{2\over{\eps}}
 \ln\left({-s\over {-t}}\right)+\ord(\eps)\,.
\label{oneloopback} \end{eqnarray}
 to leading accuracy in $s/u$. Note that
on the right hand side the first term, which is proportional to
the tree amplitude Eq.~(\ref{elas}), is real. The second term contains
the new colour structure introduced at the one loop level and  is
proportional to the tree sub-amplitude (\ref{suba}). It is purely
imaginary, since to this accuracy $\ln(-s/-t)\simeq -i\pi$, where
we used the usual prescription $\ln(-s) = \ln(s) - i\pi$, for $s >
0$. In \eqn{oneloopback} and further in our paper we have rescaled
the coupling as
\beq \tgs^2(u) = \gs^2 \cg \left({\mu^2\over -u}\right)^{-\eps}\,
.\label{rescal} \eeq

In the one-loop amplitude helicity is not conserved in the $s$
channel, thus the unequal helicity 
coefficient functions of type $C^{(1)}(\pm\mp)$
are no longer subleading. In fact, from Ref.~\cite{Kunszt:1994sd} we
obtain
\beq%
 {\cM}^{(1)}(p_a^-, p_b^{\mp};p_{a'}^-,p_{b'}^{\pm}) =%
2i{\gs^4 \over (4\pi)^2}(\CA-\CF)%
\sqrt{s\over -u}(T^{b}T^{b'})_{a'a}\, , \label{viol} \eeq
which is finite and can be written in the form of Eq.~(\ref{elas}) by
replacing the one of the helicity conserving coefficient functions of type
$C^{(0)}$ with the helicity violating coefficient function of type $C^{(1)}$
where
\beq C^{(1)}_{q\,g}(p_a^-,p_{b'}^+) = C^{(1)}_{g
q}(p_b^+,p_{a'}^-) = -{\gs^2\over (4\pi)^2}(\CA-\CF)
{|p_{a'_\perp}|\over p_{a'_\perp}}\, .\eeq
Note that in the calculation of a production rate the coefficient
function of type $C^{(1)}(\pm\mp)$ appear only in the square of a
one-loop amplitude.

Taking the interference of the one loop amplitude
(\ref{oneloopback}) with the tree amplitude, (\ref{elas}), and
summing over helicity and colour of initial and final states, we
obtain,
\bea \lefteqn{\sum_{hel,\, col} \left( {\cM}^{(1)} {\cM}^{(0)*}
\right) = \sum_{hel,\, col} |{\cM}^{(0)}|^2\ \tgs^2(u)}  \label{interf}\\%
& & \times\left[ {2C_{\sss F}\over {-\eps}} \ln\left({s\over -
u}\right) - {2(\CA + \CF )\over \eps^2} + {3\CF\over\eps} + (\pi^2
-7) \CF - {i\pi\over\eps} {1\over \CF} \right]+\ord(\eps). \nn \eea
\subsection{The colour structure  in the $u$-channel}

In order to gain more insight in how the amplitudes in the colour
decomposition (\ref{eq:coldec}) are related to the exchange of
particular colour states, we decompose the quark-gluon scattering
amplitudes according to the irreducible colour representations,
$\underline{3}\otimes\underline{8} =
\underline{3}\oplus\underline{\bar 6}\oplus\underline{15}$,
exchanged in the $u$ channel,
\beq
{\cal M} = \sum_\chi (P^{bb'})_{a'a}
(\chi)\ {\cal M}_\chi\,
,\label{projamp}
\eeq
where we recall that $a$ and $a'$ are quark colour indices and $b$ and
$b'$ are gluon colour indices, and $\chi = \underline{3},\
\underline{\bar 6},\
\underline{15}$. $M_\chi$ are colour-stripped coefficients
and $(P^{bb'})_{a'a}(\chi)$ are the colour projectors,
\bea
(P^{bb'})_{a'a}(\underline{3}) &=&
{1\over\CF} (T^b T^{b'})_{a'a}\, ,\nn\\
(P^{bb'})_{a'a}(\underline{\bar 6}) &=&
{1\over 2} \delta^{bb'} \delta_{a'a} -
{1\over N-1} (T^b T^{b'})_{a'a} - (T^{b'} T^b)_{a'a}\,
,\nn\\ (P^{bb'})_{a'a}(\underline{15}) &=&
{1\over 2} \delta^{bb'} \delta_{a'a} -
{1\over N+1} (T^b T^{b'})_{a'a} + (T^{b'} T^b)_{a'a}\,
,\label{proj}
\eea
which fulfill the usual property of projectors,
\beq
(P^{bb'})_{a'a}(\chi)\; (P^{b'b''})_{a a''}(\chi')
= \delta_{\chi\chi'} (P^{bb''})_{a'a''}(\chi)\,
.\label{ortho}
\eeq
We find that
\begin{eqnarray}
{\cal M}_{\underline{3}} &=& 2\CF {\cal A}-\frac{1}{N} {\cal B}+{\cal
C}\, ,\nonumber \\
{\cal M}_{\underline{\bar 6}}  &=&-{\cal B}+{\cal C}\, ,\nonumber \\
{\cal M}_{\underline{15}}  &=&{\cal B}+{\cal C}\, .
\end{eqnarray}

\subsection{Signature of the exchanged state}

In addition to having a particular colour, exchanged Reggeons
must have a particular signature. In other words they are either even (positive
signature) or odd (negative signature) under the exchange of $s$ and $t$.
We therefore define the amplitudes of specific signature to be,
\begin{equation}
{\cal M}^{\pm}_\chi
= {1\over 2} \left( {\cal M}_\chi\pm {\cal M}_\chi( s \leftrightarrow t)\right). \label{sigdecomp}
\end{equation}
Note that only the colour triplet positive signature exchange is
expected to Reggeize.  This is the contribution that will generate
the LL and NLL behaviour of the amplitude. The other positive
signature contributions and all of the negative signature
contributions are not given by simple poles. In the LL
approximation they can be obtained using the equation for
amplitudes with fermion exchange derived in~\cite{Fadin:1977jr},
but beyond the LL it is not known how to analyse  these
structures. Note also that ${\cal M}_\chi( s \leftrightarrow t)$
is in fact the amplitude for quark-antiquark annihilation into
gluons.

As mentioned earlier, the colour structure $\delta^{bb'} \delta_{a'a}$
occurs first at one-loop, while
the Born contribution to $(T^{b'} T^b)_{a'a}$ vanishes in the
Regge limit.
Therefore, in the Regge limit, many of the amplitudes ${\cal
M}^{(0)\pm}_{\chi}$ vanish in the Born approximation.
In fact, from \eqns{suba}{elas} we obtain,
\begin{equation}
{\cal M}^{(0)-}_{\underline{3}} ={\cal M}^{(0)+}_{\underline{\bar
6}} ={\cal M}^{(0)-}_{\underline{\bar 6}} = {\cal
M}^{(0)+}_{\underline{15}} = {\cal M}^{(0)-}_{\underline{15}} =
0\, ,
\end{equation}
and only ${\cal M}^{(0)+}_{\underline{3}}$ is non-zero,
\begin{equation}
\cM^{(0)+}_{\underline{3}} = 2\CF \mathcal{A}^{(0)}\, .
\eeq
From \eqn{oneloopback}, for the one-loop coefficient of the positive 
signature triplet, we have
\bea
\lefteqn{ \cM^{(1) +}_{\underline{3}} = 
\tgs^2(u) \cM^{(0)+}_{\underline{3}} }\nn\\
&& \times
\left\{ {C_{\sss F}\over -\eps} 
\left[ \ln\left({s\over -u}\right) + \ln\left({- s\over -u}\right) \right]
- {2(\CA + \CF )\over \eps^2} + {3\CF\over\eps} + (\pi^2 -7) \CF \right\}
+ \ord(\eps),\nn \label{posdecomp}
\eea
while for the higher-dimensional representations we have,
\beq
\cM^{(1) +}_{\underline{\bar 6}} = \cM^{(1) +}_{\underline{15}} = 0\, 
.\label{posdecomp2}
\eeq
Thus, the colour representations $\underline{\bar 6}$ and $\underline{15}$
are not present in the positive signature, at one-loop level and to 
leading power accuracy in $s/u$.  In fact,
in the LL approximation~\cite{Fadin:1977jr},
\begin{equation}
{\cal M}^{+}_{\underline{\bar 6}} = {\cal M}^{+}_{\underline{15}}
=0 \label{sign+}
\end{equation}
to all orders.

For the one-loop coefficients of negative signature, we obtain,
\bea
\cM^{(1) -}_{\underline{3}} &=& \tgs^2(u)\ \cM^{(0)+}_{\underline{3}}\
{1\over \eps}\ \left\{ - \CF
\left[ \ln\left({s\over -u}\right) - \ln\left({- s\over -u}\right) \right]
+ {1\over\CF} \ln\left({-s\over -t}\right) \right\} + \ord(\eps) \, ,\nn\\
\cM^{(1) -}_{\underline{\bar 6}} &=& \cM^{(1) -}_{\underline{15}} = 
\tgs^2(u)\ \cM^{(0)+}_{\underline{3}}\ {1\over \eps}\ {1\over\CF}
\ln\left({-s\over -t}\right) + \ord(\eps)\, .\label{negdecomp}
\eea
Thus in the one-loop amplitude of negative signature all the three
representations exchanged in the $u$ channel contribute, to leading 
power accuracy in $s/u$.

\section{Regge Theory Interpretation}

Let's choose, for definiteness, the
QCD Compton scattering process
\begin{equation}
q(p_a)+ g(p_b)\rightarrow q(p_{a'}) +g(p_{b'}).
\end{equation}
We will use physical polarizations of gluons, so that their
polarization vectors satisfy $e(p_b)\cdot p_b=e(p_b)\cdot
p_{b'}=0$ and $e(p_{b'})\cdot p_{b'}=e(p_{b'})\cdot p_b=0$. Then,
for massless quarks, the contribution of the Reggeized quark to
the amplitude for colour triplet exchange with even signature can
be presented~\cite{Fadin:1977jr} as,
\begin{equation}
{\cal R}_{\underline{3}}^+=
\Gamma_{QG}\frac{-1}{\not q_{\perp}}
\frac{1}{2}\left[
\left( \frac{-s}{-u}\right) ^{\delta_T(u)}
+\left(\frac{s}{-u}\right) ^{\delta_T(u)}\right]
\Gamma_{GQ}~, \label{z1}
\end{equation}
where $q_{\perp}$ is the transverse to the $(p_a,p_b)$ plane part
of the momentum transfer $q=p_{a'}-p_b$, $u=q^2=q_\perp^2$ in the
Regge limit, $\Gamma_{QG}$ and $\Gamma_{GQ}$ are the Reggeon
vertices for the $G\rightarrow Q$ and $Q\rightarrow G$ transitions
and $\delta_T$ determines the quark Regge trajectory.   Note that
for massless quarks $\delta_T$ depends only on $q_{\perp}^2$, so
that instead of two complex conjugate trajectories with opposite
parities for massive quarks we have a single one. We assume that
the quark Regge trajectory has the perturbative expansion,
\begin{equation}
\delta_T(u) = \tgs^2(u) \; \delta_T^{(1)}
+\tgs^4(u) \; \delta_T^{(2)}  + \ord(\tgs^6(u))\,.
\end{equation}
At leading
order~\cite{Fadin:1977jr} and in $D=4+2\eps$ dimensions,
\begin{equation}
\delta^{(1)}_T= -{2\CF\over \eps}.
\end{equation}
This is related to the analogous one-loop gluon trajectory by
$$
\delta^{(1)}_T = {\CF\over \CA} \omega^{(1)}.
$$

The general structure of the Reggeon vertices is determined
by relativistic invariance and colour symmetry.  For the quark-gluon
vertices the general structure is
\begin{eqnarray}
\Gamma_{QG}&=&-\gs\bar{u}(p_{a'})T^{b}\left[\not{\!e(p_{b})}(1+\delta_e(u))
+\frac{e(p_{b})\cdot q\not{\!q}_\perp}{{q}^2_\perp}\delta_q(u)\right]~,\nonumber \\
\Gamma_{G Q}&=&-\gs\left[\not{\!e^*(p_{b'})}(1+\delta_e(u))
+\frac{e^*(p_{b'})\cdot q\not{\!q}_\perp}{{q}^2_\perp}\delta_q(u)\right]T^{b'}
u(p_{a})~.  \label{z3}
\end{eqnarray}
Note that it is straightforward to relate this general vertex structure to the
scattering of particular particle helicities, \sec{sec:twoone}.
The functions $\delta_e$
and $\delta_q$ represent the radiative corrections to the Born vertices.  They
are functions of $q_\perp ^2$ in the massless case and
have the perturbative expansion
\begin{eqnarray}
\delta_e(u) &=& \tgs^2(u)\; \delta^{(1)}_e
+\tgs^4(u)\; \delta^{(2)}_e + \ord(\tgs^6)\,,\nonumber \\
\delta_q(u) &=& \tgs^2(u)\; \delta^{(1)}_q
+\tgs^4(u)\;\delta^{(2)}_q + \ord(\tgs^6)\,.
\end{eqnarray}
In the one-loop approximation the corrections were obtained in
\cite{Fadin:2001dc} and have the form
\begin{eqnarray}
\delta_e^{(1)}
&=& \omega^{(1)}\left[\frac{\CF}{2\CA}\left(\frac{1}{\epsilon}-
\frac{3(1-\epsilon)}{2(1+2\epsilon)}+\psi(1)+\psi(1-\epsilon)-
2\psi(1+\epsilon)\right)+\frac{1}{2\epsilon}-\frac{\epsilon}{2(1+2\epsilon)}\right
]~,\nonumber \\
&&
\label{z4}
\\
\delta_q^{(1)} &=&\omega^{(1)}\frac{\epsilon}{2(1+2\epsilon)}\left(1+
\frac{1}{N^2}\right)~. \label{z5}
\end{eqnarray}
Note,  that whereas $\delta_e^{(1)}$ has a soft ($1/\epsilon^2$) singularity that must
be cancelled by real radiation,
$\delta_q^{(1)}$ is finite as $\epsilon \rightarrow 0$,
since the corresponding spin structure  is absent at leading
order.

There is no ansatz for any of the odd signature exchanges or for
the even signature  $\underline{\bar 6}$ and $\underline{15}$
exchanges. These contributions do not correspond to simple poles
and cannot be described by an ansatz of the form of \eqn{z1}. As
mentioned earlier, in  the LL approximation ${\cal
M}^{+}_{\underline{\bar 6}}$ and ${\cal M}^{+}_{\underline{15}}$
are zero to all orders. The LL
negative signature contributions
through to ${\cal O}(D-4)$ 
can be obtained from the equation derived
in~\cite{Fadin:1977jr} and are given by
 \beq
 {\cM}_{\underline{3}}^-
 ={\cM}_{\underline{3}}^{(0)+}\tgs^2(u)\left(\CF+\frac{1}{\CF}\right)
 \frac{1}{\eps}\left(-i\pi +\tgs^2(u)\frac{\CF}{\eps}(2i\pi\ln\left(\frac{s}{|u|}\right)+\pi^2 )+
 \ord(\tgs^4(u))\right),
 \label{M3-}
 \eeq
 \beq
 {\cM}_{\underline{\bar 6 }}^-
 ={\cM}_{\underline{3}}^{(0)+}\tgs^2(u)\frac{1}{\CF\eps}\left(-i\pi
 +\tgs^2(u)\frac{\CA+2\CF-1}{2\eps}(2i\pi\ln\left(\frac{s}{|u|}\right)+\pi^2 )+
 \ord(\tgs^4(u))\right),
 \label{M6-}
 \eeq
 \beq
 {\cM}_{\underline{15 }}^-
 ={\cM}_{\underline{3}}^{(0)+}\tgs^2(u)\frac{1}{\CF\eps}\left(-i\pi
 +\tgs^2(u)\frac{\CA+2\CF+1}{2\eps}(2i\pi\ln\left(\frac{s}{|u|}\right)+\pi^2)+
 \ord(\tgs^4(u)) \right).
 \label{M15-}
 \eeq
Note that at the one-loop level Eqs. (\ref{sign+}), (\ref{z1})-(\ref{z5}) and
(\ref{M3-})-(\ref{M15-}) give all of the contributions to the scattering
amplitude that survive in the Regge limit. They are in agreement with 
Eqs.~(\ref{oneloopback}), (\ref{viol}), (\ref{posdecomp}), (\ref{posdecomp2}) 
and (\ref{negdecomp}) obtained from exact calculations.

\subsection{Projection by tree-level amplitude}

Let us denote the projection of the tree amplitude ${\cal M}^{(0)}$ summed over
spins and colours on a generic amplitude ${\cal M}$ as,
\begin{equation}
\langle {\cal M}^{(0)}| {\cal M}\rangle = \sum_{spin,\;col} {\cal
M}^{(0)^\dagger} {\cal M}.
\end{equation}

For the Reggeized amplitude we obtain the projection normalised by the square of
the Born
amplitude to be,
\begin{eqnarray}
\lefteqn{ \frac{\langle{\cal M}^{(0)}|{\cal R}^+_{\underline{3}}\rangle}
{\langle {\cal M}^{(0)}|{\cal M}^{(0)}\rangle} }\label{z6}\\
&=& \exp{(\delta_T(u)\, L)}
\frac{\left(1+\exp^{-i\pi\delta_T(u)}\right)}{2}
\left[(1+\delta_e(u))^2+
\frac{(1+\delta_e(u))\delta_q(u)}{1+\epsilon}
+\frac{\delta_q^2(u)}{4(1+\epsilon)^2}\right]\, ,\nn
\end{eqnarray}
where $L=\ln(s/-u)$.
Note that ${\cal R}^{(0)+}_{\underline{3}}$ coincides with the Born amplitude
so that setting $\delta_e = \delta_T = \delta_q = 0$ produces unity.

Writing ${\cal R}^+_{\underline{3}}$ as a perturbative series,
\begin{equation}
{\cal R}^+_{\underline{3}}  =  \sum_n
 \tgs^{2n}(u) \;
{\cal R}^{(n)+}_{\underline{3}},
\end{equation}
and expanding \eqn{z6}  to first order
gives the one-loop contribution to the Reggeized amplitude ${\cal R}^{(1)+}_{\underline{3}}$ such that
\begin{equation}
\frac{\langle{\cal M}^{(0)}|{\cal R}^{(1)+}_{\underline{3}}\rangle}
{\langle {\cal M}^{(0)}|{\cal M}^{(0)}\rangle}
=
\delta^{(1)}_T\,L+2\delta_e^{(1)}+\frac{\delta_q^{(1)}}{1+\epsilon}
-i{\pi\over 2} \delta^{(1)}_T\,,
\label{z7}
\end{equation}
that agrees with the results of \cite{Ellis:1986er,Kunszt:1994sd}.
For the two-loop contribution we obtain
\begin{eqnarray}
\frac{\langle{\cal M}^{(0)}|{\cal R}^{(2)+}_{\underline{3}}\rangle}
{\langle {\cal M}^{(0)}{\cal M}^{(0)}\rangle}
&=&
\frac{(\delta^{(1)_T})^2}{2}\,L^2+
\left[\delta^{(2)}_T+\left(2\delta_e^{(1)}+\frac{\delta_q^{(1)}}{1+\epsilon}
\right)\delta^{(1)}_T\right]L
\nonumber \\
&&-\frac{\pi^2}{4}(\delta^{(1)}_T)^2+(\delta_e^{(1)})^2+
\frac{\delta_e^{(1)}\delta_q^{(1)}}{1+\epsilon} +
\frac{(\delta_q^{(1)})^2}{4(1+\epsilon)^2}+2\delta_e^{(2)}
+\frac{\delta_q^{(2)}}{1+\epsilon}\,\nonumber \\
&& - i\frac{\pi}{2}\left( (\delta^{(1)}_T)^2\,L
+ \frac{\delta_q^{(1)}\delta_T^{(1)}}{1+\eps}
+ 2\delta_e^{(1)} \delta_T^{(1)}
+\delta_T^{(2)}\right).
\label{z8}
\end{eqnarray}

Note that the Reggeized form for positive signature colour triplet exchange
does not saturate the possible contributions to the cross section in the Regge
limit.  There is also a contribution from the negative signature exchange.
This negative signature contribution appears first in the imaginary part of the
(normalized)  projection on the one-loop amplitude  and also in the logarithmic
imaginary part and non-logarithmic real parts of the two-loop amplitude.

Therefore, if the statement about the quark
Reggeization is valid in the next-to-leading logarithmic order,
the logarithmic terms in the
right-hand side of \eqns{z7}{z8} must coincide with
corresponding terms for the
total amplitude, which can be found using the
results of \cite{Anastasiou:2001sv} in the appropriate limit.

\section{The Regge limit of the one- and two-loop calculations}

The interference of the tree and one-loop amplitudes for quark-gluon scattering
has been given in \cite{Ellis:1986er} while the interference of the tree and two-loop
amplitudes has been computed in \cite{Anastasiou:2001sv}.
Taking the Regge limit or leading
power of $s/|u|$ of these
unrenormalised expansions, then we write,
\begin{eqnarray}
\frac{{\rm Re}\langle{\cal M}^{(0)}|{\cal M}^{(n)\pm}_{\underline{3}}\rangle}
{\langle \cM^{(0)}|\cM^{(0)}\rangle}
&=& \tgs^{2n}(u)
\sum_{m=0}^n B_{nm}^\pm \ln^m\left({s\over - u}\right) ,\label{eq:expandpm}\\
\frac{{\rm Im}\langle{\cal M}^{(0)}|{\cal
M}^{(n)\pm}_{\underline{3}}\rangle} {\langle
\cM^{(0)}|\cM^{(0)}\rangle} &=& -\frac{\pi}{2} \tgs^{2n}(u)
\sum_{m=0}^{n-1} D_{nm}^\pm \ln^m\left({s\over - u}\right)
,\label{eq:expandimpm}
\end{eqnarray}
where the positive and negative signature pieces are constructed according to
\eqn{sigdecomp}.
For $n=0$, $B_{00}^+ =1$ and $B_{00}^- =0$. In \eqn{eq:expandimpm}, $n\ge 1$.
In both \eqns{eq:expandpm}{eq:expandimpm} a sum over colours and helicities
is implicit.
For the interference of tree with one-loop, the coefficients of the
positive signature amplitudes are,
\bea
B^{+}_{11} &=& -\frac{2}{\eps}\CF\ , \nn \\
B^{+}_{10} &=& 2\CA \left(-{1\over \eps^2} +\eps-3 \eps^2\right) \nn \\
&+& 2\CF \left(-{1\over \eps^2} + {3\over 2 \eps}+ {(\pi^2-7)\over 2} -
(\zeta_3-6)~\eps + {(\pi^4-330)\over 30}~\eps^2
\right)\ ,\nn \\  \nn \\
D^{+}_{10} &=& -\frac{2\CF}{\eps}\ ,\label{eq:b11plus}
\eea
and those of negative signature are
\bea
B^{-}_{11} &=& 0\ ,  \nn \\
B^{-}_{10} &=& 0\ ,  \nn \\
D^{-}_{10} &=& +\frac{2}{\eps \CF} + \frac{2\CF}{\eps}\ .\label{eq:b11minus}
\eea
We immediately see that all of the positive signature contributions agree
with the expansion of \eqn{z7} about $\eps = 0$.  The leading and
next-to-leading logarithmic terms $B^+_{11}$ and $B^+_{10}$ precisely match up
while the corresponding  negative signature terms are not present.
The negative signature contribution is purely imaginary in this order.
It is determined by the LL approximation and is in accordance with
Eq.~(\ref{M3-}).

At the two-loop level, the amplitude for quark-gluon scattering is not
known as such, but it has been computed at the level of the interference
with the tree amplitude~\cite{Anastasiou:2001sv}, using conventional
dimensional regularization (CDR) and renormalised in the \MS\ scheme.
In Ref.~\cite{Anastasiou:2001sv} the divergent contribution is written in
terms of the infrared singularity operators $\ione$, $\itwo$ and $\htwo$
introduced by Catani~\cite{Catani:1998bh}
and the tree- and one-loop amplitudes.
The
finite remainder is given in terms of logarithms and polylogarithms with
arguments $-u/s$, $-t/s$ and $u/t$.
Making the same expansion in the high energy limit and keeping only the leading
power of $s/|u|$, we can extract the two-loop coefficients  of the
positive signature amplitudes,
\bea
B^+_{22} &=& {1\over 2} \left( B^+_{11}\right)^2\ ,\label{eq:b22plus}\\
B^+_{21} &=& B^+_{11} B^+_{10} + \CF \beta_0 {2\over \eps^2} - \CF
K {2 \over \eps}\nn\\
&+&\CF\CA \left({404\over 27} -2 \zeta_3\right) + \CF\NF
\left(-{56\over 27}\right)\nn\\
&+&\CF(\CF-\CA) \left(16\zeta_3\right)\ , \label{eq:b21plus} \\
B^+_{20} &=& {1\over 2} \left( B^+_{10}\right)^2 + \CA\beta_0
{1\over \eps^3} + \CF\beta_0 {1\over \eps^3}
\nn\\
&+& \CA^2 \Biggl(\left(-{67\over 18} +{\pi^2\over 6}\right){1\over
\eps^2}
-\left( -{193\over 27} + {11\pi^2\over 18} + \zeta_3 \right){1\over \eps}
\nn \\
&+&\left(-{1736\over 81}+{67\pi^2\over 54} + 4 \zeta_3
+{\pi^4\over 12}
\right)
\Biggr) \nn \\
&+& \CA\CF \Biggl( \left(-{83\over 9} +{\pi^2\over 6}\right){1\over
\eps^2}
-\left(-{3733\over 108} +{55\pi^2\over 18} + 13\zeta_3\right){1\over \eps}
\nn \\
&+&\left(-{71929\over 648} + {353\pi^2\over 54} + {242\over 3}\zeta_3
+ {7\pi^4\over 60}
\right)
\Biggr) \nn \\
&+& \CF^2 \left( -\left(-{3\over 4} +\pi^2 -12 \zeta_3\right){1\over
\eps}+\left(-{9\over 8} + 2\pi^2 - 42\zeta_3 -{28\pi^4\over
45}\right)
\right) \nn \\
&+& \CA\NF \left( {5\over 9\eps^2} - \left({19\over 27} - {\pi^2 \over
9}\right){1\over \eps}
+\left({173\over 81} - {5\pi^2\over 27} -4\zeta_3\right)
\right) \nn \\
&+& \CF\NF \left({14\over 9\eps^2} - \left({317\over 54} - {5\pi^2 \over
9}\right){1\over \eps}
+\left( {5993\over 324} -{25\pi^2\over 27} - {8\over 3}\zeta_3\right)
\right) \nn \\
&+& {\CA\over \CF} \left(-{\pi^2\over\eps^2} \right)\nn \\
&-& (\CF-\CA)^2~{\pi^2\over \eps^2} \, ,
 \label{eq:b20plus}
\eea
\bea
D^{+}_{21} &=& \left(B^+_{11}\right)^2\label{eq:d21plus} \, ,\\
D^{+}_{20} &=& B^+_{21}\ ,\label{eq:d20plus}
\eea
where
\beq
\beta_0= {(11\CA-2N_F)\over 6}, \qquad\qquad K = \left({67\over 18} -
{\pi^2\over 6} \right) \CA - {5\over 9} N_F\, .\label{eq:beta}
\eeq
Comparing the leading logarithmic contribution $B^+_{22}$ (and $D^+_{21}$)
with \eqn{z8}
we immediately see that it is exactly as predicted - and is nothing more than
a confirmation of the exponentiation of the leading logarithms.  The
next-to-leading logarithmic term $B^+_{21}$ contains two pieces - one which is
an echo of the one-loop coefficient function and the one-loop quark trajectory
$B^+_{10}B^+_{11}$, and the other is the two-loop quark trajectory $\delta_T^{(2)}$ which
appears for the first time,
\beq
\delta^{(2)}_T = \CF \left[ \beta_0 {2\over \eps^2} - K {2
\over \eps} + \CA \left({404\over 27} -2 \zeta_3\right)
+ \NF \left(-{56\over 27}\right)
+ (\CF-\CA) \left(16\zeta_3\right) \right]\, .\label{twolooptraja}
\eeq
Note that \eqn{twolooptraja} has the remarkable feature that by mapping
$\CF \to \CA$, we obtain the two-loop gluon Regge trajectory.
The full quark trajectory through to two-loop order is thus,
\begin{eqnarray}
j_Q(u)&=&{1\over 2}+\tgs^2(u)\;\frac{\omega^{(1)}}{N}\CF
\left[1+\tgs^2(u)\;\frac{\omega^{(1)}}{2N}
\biggl\{\beta_0-K\epsilon
\right.\nn \\
&&\hspace{3cm}+\left.\left.\left(\left(\frac{202}{27}-9\zeta(3)\right) N
-\frac{28}{27}N_F+8\zeta(3) \CF \right)\epsilon^2
\right\}\right]\, .  \label{twolooptrajb}
\end{eqnarray}

The non-logarithmic term in the real part, $B^+_{20}$, belongs to the
next-to-next-to-leading logarirhmic contribution and is not
expected to match up with the ansatz of \eqn{z1} because of
possible Regge cut contribution; conversely, the non-logarithmic term in the
imaginary part, $D^+_{20}$, is next-to-leading and should be given by the
anzatz of \eqn{z1}. Comparing to \eqn{z8} one can  easily see
that it is the case.

Similarly the negative signature coefficients are given by
\bea
B^{-}_{22} &=& 0\ , \nn \\
B^{-}_{21} &=& 0\ , \nn \\
B^{-}_{20} &=& (\CF-\CA)^2~{\pi^2\over \eps^2}\ , \label{eq:b2minus} \\
D^{-}_{21} &=& -{4 \over \eps^2} - {4\CF^2\over \eps^2}\ ,\nn \\
D^{-}_{20} &=& -\left(1+{1\over \CF^2}\right) B^+_{21}
+ {\CA\over \CF} \left(10-4\pi^2-8\zeta_3\right)
+{\NF\over \CF} 12\nn \\
&+& \CF^2 \left(-8 +8\pi^2+16\zeta_3\right)
+\CA^2\left(-20+8\pi^2+16\zeta_3\right)\nn\\
&
+&\CA\CF \left(44-20\pi^2-40\zeta_3\right)
+72\CF\NF-36 \CA\NF\ .\label{eq:d2minus}
\eea
The coefficients  $B^-_{22}$ and $B^-_{21}$ vanish simply due to signature;
$B^-_{20}$ and  $D^-_{21}$ are determined by the LL approximation and are 
in accordance with Eq.~(\ref{M3-}). It is not presently known how to interpret the coefficient $D^-_{20}$.

\section{Conclusions}

By taking the high-energy limit of the two-loop amplitudes for
quark-gluon scattering~\cite{Anastasiou:2001sv}, we have tested
the validity of the general form of the high-energy  amplitudes
(\ref{z1}) for quark-gluon scattering, arising from a Reggeized
quark with colour triplet and even signature exchanged in the
crossed channel. The limit is compatible with the Reggeization in
the leading and the next-to-leading orders.  We have therefore
extracted two-loop Regge trajectory for the quark,
\eqns{twolooptraja}{twolooptrajb}, as an expansion in powers of
$(D-4)$ for the space-time dimension $D$ tending to the physical
value $D=4$.  At present it is not known how to describe either
the next-to-next-to-leading logarithmic triplet exchange contributions 
for either the positive signature $B^+_{20}$
(given in \eqn{eq:b20plus}) or the  negative signature $D^-_{20}$ 
(given  in \eqn{eq:d2minus}).

\section*{Acknowledgements}

We thank Stefano Catani and Zoltan Trocsanyi for helpful comments.
VDD thanks the IPPP and the CERN Theory Division for their kind
hospitality during the early stage of this work. This work was
supported in part by the EU Fourth Framework Programme `Training
and Mobility of Researchers', Network `Quantum Chromodynamics and
the Deep Structure of Elementary Particles', contract
FMRX-CT98-0194 (DG-12-MIHT), and grants INTAS 97-31696, 00-00366
and RFBR  01-02-16042.


\begin{thebibliography}{999}

\bibitem{gellmann}
M. Gell-Mann, M.L. Goldberger, F.E. Low, E. Marx and F. Zachariasen,
{\it Elementary Particles of Conventional Field Theory as Regge Poles. III},
{\sl Phys. Rev.} {\bf 133} (1964) B145.

\bibitem{McCoy:1976ff}
B.~M.~McCoy and T.~T.~Wu,
{\it Theory Of Fermion Exchange In Massive Quantum Electrodynamics
At High-Energy. I,}
\prd{13}{1976}{369}.

\bibitem{mandel}
S.~Mandelstam,
{\it Non-Regge Terms in the Vector-Spinor Theory},
{\sl Phys. Rev.} {\bf 137} (1965) B949.



\bibitem{Grisaru:1973vw}
M.~T.~Grisaru, H.~J.~Schnitzer and H.-S.~Tsao,
{\it Reggeization of yang-mills gauge mesons in theories with a
                  spontaneously broken symmetry,}
\prl{30}{1973}{811}.

\bibitem{Grisaru:1973cf}
M.~T.~Grisaru and  H.~J.~Schnitzer,
{\it Reggeization of elementary particles in renormalizable
                  gauge theories - vectors and spinors,}
\prd{8}{1973}{4498}.

\bibitem{Lipatov:1976zz}
L.~N.~Lipatov,
{\it Reggeization Of The Vector Meson And The Vacuum Singularity In
Nonabelian Gauge Theories,}
\yf{23}{1976}{642} [\sjnp{23}{1976}{338}].

\bibitem{Fadin:1975cb}
V.~S.~Fadin, E.~A.~Kuraev and L.~N.~Lipatov,
{\it On the Pomeranchuk singularity in asymptotically free theories,}
\plb{60}{1975}{50}.

\bibitem{Kuraev:1976ge}
E.~A.~Kuraev, L.~N.~Lipatov and V.~S.~Fadin,
{\it Multi - Reggeon Processes In The Yang-Mills Theory,}
\zetf{71}{1976}{840} [\jetp{44}{1976}{443}].

\bibitem{Kuraev:1977fs}
E.~A.~Kuraev, L.~N.~Lipatov and V.~S.~Fadin,
{\it The Pomeranchuk Singularity In Nonabelian Gauge Theories,}
\zetf{72}{1977}{377} [\jetp{45}{1977}{199}].

\bibitem{Balitsky:1978ic}
I.~I.~Balitsky and L.~N.~Lipatov,
{\it The Pomeranchuk Singularity In Quantum Chromodynamics,}
\yf{28}{1978}{1597} [\sjnp{28}{1978}{822}].

\bibitem{Fadin:1977jr}
V.~S.~Fadin and V.~E.~Sherman, {\it Fermion Reggeization in
non-Abelian gauge theories, } Pis'ma Zh.\ Eksp.\ Teor.\ Fiz.\
{\bf 23} (1976) 599; {\it Processes with fermion exchange in
nonabelian gauge theories,} \zetf{72}{1977}{1640}
[\sjnp{45}{1977}{861}].



\bibitem{Fadin:1998fv}
V.~S.~Fadin and R.~Fiore,
{\it The generalized nonforward BFKL equation and the
                  bootstrap condition  for the gluon Reggeization in the
                  NLLA,}
\plb{440}{1998}{359} [hep-ph/9807472].

\bibitem{Fadin:1998py}
V.~S.~Fadin and L.~N.~Lipatov,
{\it BFKL pomeron in the next-to-leading approximation,}
\plb{429}{1998}{127} [{\tt hep-ph/9802290}].

\bibitem{Camici:1997ij}
G.~Camici and M.~Ciafaloni,
{\it Irreducible part of the next-to-leading BFKL kernel,}
\plb{412}{1997}{396};
Erratum \plb{417}{1998}{390} [{\tt hep-ph/9707390}].

\bibitem{Ciafaloni:1998gs}
M.~Ciafaloni and G.~Camici,
{\it Energy scale(s) and next-to-leading BFKL equation,}
\plb{430}{1998}{349} [{\tt hep-ph/9803389}].

\bibitem{Fadin:1995xg}
V.~S.~Fadin, M.~I.~Kotsky and R.~Fiore,
{\it Gluon Reggeization in QCD in the next-to-leading order,}
\plb{359}{1995}{181}.

\bibitem{Fadin:1996tb}
V.~S.~Fadin, R.~Fiore and M.~I.~Kotsky,
{\it Gluon Regge trajectory in the two-loop approximation,}
\plb{387}{1996}{593} [{\tt hep-ph/9605357}].

\bibitem{Fadin:1996km}
V.~Fadin, R.~Fiore and A.~Quartarolo,
{\it Reggeization of quark-quark scattering amplitude in QCD,}
\prd{53}{1996}{2729} [{\tt hep-ph/9506432}].

\bibitem{Blumlein:1998ib}
J.~Blumlein, V.~Ravindran and W.~L.~van Neerven,
{\it On the gluon Regge trajectory in $O(\alpha_s^2)$,}
\prd{58}{1998}{091502} [{\tt hep-ph/9806357}].

\bibitem{DelDuca:2001gu}
V.~Del Duca and E.~W.~N.~Glover,
{\it The high energy limit of QCD at two loops,}
{\it JHEP} {\bf 0110} (2001) 035
[{\tt hep-ph/0109028}].


\bibitem{Ellis:1986er}
R.~K.~Ellis and J.~C.~Sexton,
{\it QCD Radiative Corrections To Parton Parton Scattering,}
\npb{269}{1986}{445}.

\bibitem{Kunszt:1994sd}
Z.~Kunszt, A.~Signer and Z.~Trocsanyi,
{\it One loop helicity amplitudes for all 2 $\to$ 2 processes in QCD and
N=1 supersymmetric Yang-Mills theory,}
\npb{411}{1994}{397} [\hepph{9305239}].

\bibitem{Anastasiou:2001sv}
C.~Anastasiou, E.~W.~N.~Glover, C.~Oleari and M.~E.~Tejeda-Yeomans,
{\it Two-loop QCD corrections to massless quark gluon scattering,}
\npb{605}{2001}{486} [\hepph{0101304}].

\bibitem{Mangano:1991by}
M.~L.~Mangano and S.~J.~Parke, {\it Multiparton amplitudes in
gauge theories,}
\prep{200}{1991}{301}.

\bibitem{Parke:1986gb}
S.~J.~Parke and T.~R.~Taylor, {\it An Amplitude For N Gluon
Scattering,}
\prl{56}{1986}{2459}.

\bibitem{DelDuca:2000ha}
V.~Del Duca, A.~Frizzo and F.~Maltoni, {\it Factorization of tree
QCD amplitudes in the high-energy limit and in the collinear
limit,}
\npb{568}{2000}{211} [\hepph{9909464}].

\bibitem{Fadin:2001dc}
V.~S.~Fadin and R.~Fiore,
{\it Calculation of Reggeon vertices in QCD,}
\prd{64}{2001}{114012} [\hepph{0107010}].

\bibitem{Catani:1998bh}
S.~Catani,
{\it The singular behaviour of {QCD} amplitudes at two-loop order,}
\plb{427}{1998}{161} [{\tt hep-ph/9802439}].

\end{thebibliography}
\end{document}